\documentclass[aps,superscriptaddress,,twocolumn,showpacs]{revtex4-1}
\pdfoutput=1
\usepackage{color}
\usepackage{amsmath, amsthm, amsfonts}    
\usepackage{amssymb}
\usepackage{mathptmx} 
\usepackage{dsfont}
\usepackage{cancel}
\usepackage{tikz}
\usepackage[unicode=true,pdfusetitle,
 bookmarks=true,bookmarksnumbered=false,bookmarksopen=false,
 breaklinks=true,pdfborder={0 0 0},backref=false,colorlinks=true,citecolor=blue]{hyperref}
\usepackage{graphicx}   
\usepackage[caption=false]{subfig}       
\usepackage{bm}            
\usepackage[normalem]{ulem}  

\newlength\imagewidth
\newlength\imagescale
\def\be{\begin{eqnarray}}
\def\ee{\end{eqnarray}}

\def\E{{\bf E}}
\def\H{{\bf H}}

\def\im{{\rm i}}

\definecolor{JOT-color}{named}{blue}
\definecolor{CSF-color}{named}{orange}

\begin{document}

\title{Multiple Kerker anapoles in dielectric microspheres}

\author{Cristina Sanz-Fern\'andez}
\email{cristina$_$sanz001@ehu.eus}
\affiliation{Centro de F\'{i}sica de Materiales (CFM-MPC), Centro Mixto CSIC-UPV/EHU,  20018 Donostia-San Sebasti\'{a}n, Spain}
 
\author{Mart\'in Molezuelas}
\affiliation{Centro de F\'{i}sica de Materiales (CFM-MPC), Centro Mixto CSIC-UPV/EHU,  20018 Donostia-San Sebasti\'{a}n, Spain}

\author{Jon Lasa-Alonso}
\affiliation{Centro de F\'{i}sica de Materiales (CFM-MPC), Centro Mixto CSIC-UPV/EHU,  20018 Donostia-San Sebasti\'{a}n, Spain}
\affiliation{Donostia International Physics Center (DIPC),  20018 Donostia-San Sebasti\'{a}n,  Spain}

\author{Nuno de Sousa}
\affiliation{Donostia International Physics Center (DIPC),  20018 Donostia-San Sebasti\'{a}n,  Spain}

\author{Xavier Zambrana-Puyalto}
\affiliation{Istituto Italiano di Tecnologia, Via Morego 30, 16163 Genova, Italy.}

\author{Jorge Olmos-Trigo}
\email{jolmostrigo@gmail.com}
\affiliation{Donostia International Physics Center (DIPC),  20018 Donostia-San Sebasti\'{a}n,  Spain}

\begin{abstract}
High refractive index dielectric spheres present remarkable light-scattering properties in the spectral range dominated by dipolar modes. However, most of these properties are absent for larger spheres under plane wave illumination. Here, we propose to unravel dipolar regimes regardless of the sphere size and refractive index by illuminating with a pure dipolar field. This type of illumination ensures that the scattering response of the sphere is purely dipolar. In this scenario, we show that Kerker conditions are not only related to duality symmetry and a strong backward-to-forward asymmetric light-scattering, but also to the appearance of non-radiating sources: the so-called hybrid anapoles. Finally, we show that all the abovementioned scattering features
under dipolar illumination are reproducible with an experimentally accessible tightly-focused Gaussian beam.  
\end{abstract}

\maketitle

High refractive index (HRI)  particles have received increasing interest during the past decade as building blocks of all-dielectric metamaterials and optical devices~\cite{garcia2011strong,kuznetsov2016optically}. Their unique optical properties are mostly linked with the excitation of single dipolar modes.  In contrast,  for microsized objects, these are hindered due to the contribution from higher multipolar orders under a plane wave (PW) illumination~\cite{barreda2019recent}.

The optical properties of HRI subwavelength objects include strong backward-to-forward asymmetric scattering~\cite{nieto2011angle} and, at the first Kerker condition~\cite{kerker1983electromagnetic,nieto2011angle}, where the dipolar electric and magnetic polarizabilities are identical, the emergence of the zero optical backscattering condition for nanospheres~\cite{geffrin2012magnetic}. The first Kerker condition also leads to the conservation of the incoming helicity~\cite{calkin1965invariance} and, hence, to the restoration of a non-geometrical symmetry of the electromagnetic field: duality symmetry~\cite{fernandez2013electromagnetic}. 
In the dipolar regime, the zero optical backscattering and duality restoration are linked and fully determined by the asymmetry parameter ($g$-parameter)~\cite{hulst1957light}, which describes the optical response of the particle via the interference of the electric and magnetic polarizabilities. At the first Kerker condition, the $g$-parameter is maximized ($g=0.5$), and helicity conservation leads to the absence of backscattered light~\cite{fernandez2013electromagnetic, zambrana2013duality}. 
The $g$-parameter may also be used as a signature of pure magnetic HRI dipolar scatterers, as these would necessarily lead to $g=0$.
In that scenario, it is possible to collect
only magnetic scattered light, or even no light at all. In the latter case, it is still possible to have internal electromagnetic energy via induced electric dipolar currents inside the particle~\cite{zel1958electromagnetic,  alaee2018electromagnetic, colom2019modal}. These particular optical responses, often referred to as anapoles~\cite{zel1958electromagnetic}, have attracted attention in a wide range of different fields of physics~\cite{wood1997measurement, ho2013anapole}. 
However, it is still challenging to obtain anapoles for the most studied phenomenon in the scattering of light: a homogeneous sphere under PW illumination. This difficulty arises as a result of the inevitable contribution of higher multipolar orders in the scattering of a sphere. Nevertheless, several efforts have been made in this direction, and anapoles have been observed for other geometries such as HRI nanodisks~\cite{miroshnichenko2015nonradiating, liu2017high, grinblat2017efficient, timofeeva2018anapoles, verre2019transition, yang2018anapole,savinov2019optical}, nanowires~\cite{liu2015invisible, gongora2017anapole, grinblat2020efficient, zhang2020anapole}, or core-shells~\citep{liu2015toroidal, feng2017ideal, li2020excellent}  under PW illumination. Further, tightly-focused radially polarized beams, which do not excite the magnetic
multipole components,  have been presented as a possible approach to unveil anapoles in  HRI  spheres in the limit of small particle~\cite{wei2016excitation, parker2020excitation}. However, the simultaneous suppression of the electric and magnetic dipolar scattering efficiencies for larger spheres, at the  hybrid anapole~\cite{luk2017hybrid}, is still a matter of research. 

In this Letter, we show that dipolar spectral regimes can be induced in dielectric spheres, regardless of the size parameter $x=ka$ and refractive index contrast ${{\rm{m}}}$, by illuminating with a pure dipolar field (PDF)~\cite{olmos2019sectoral}. This illumination does not excite higher multipolar orders, and hence, in analogy to HRI spheres in the limit of small particle, the scattering can be tuned from almost zero forward to perfect zero backscattering~\cite{zambrana2013dual, olmos2020optimal}. An intrinsic advantage of this type of illumination is that it allows us to find these light-scattering features for different size parameters, associated with multiple Kerker conditions. Also, we unravel the so-called hybrid anapole for several microsized dielectric spheres behaving dipolarly, unveiling a hidden connection with Kerker's conditions.  Finally, we show that experimentally accessible tightly-focused Gaussian beams (GB) mimic the abovementioned light-scattering properties under PDF illumination, which strongly motivates a future experimental verification.

The selective excitation of electric and magnetic dipolar modes by PDF has been shown in previous works~\cite{zaza2019size, huang2020surface, aibara2020dynamic} and can be explicitly derived from the multipole expansion of a circularly polarized PW,
\be   \label{plane_wave}
 \frac{ \E^{(\rm{PW})}_{\sigma}}{E_0} &=& \frac{\bm{\hat{x}} +  \im \sigma \bm{\hat{y}}}{\sqrt{2}}  e^{\im kz} = \sum_{\ell=0}^\infty \sum_{m_z=-\ell}^{+\ell} \sum_{\sigma' = \pm1} C_{\ell m_z}^{\sigma \sigma'} \bm{\Psi}^{ \sigma'}_{\ell m_z},
\ee
where  $C_{\ell m_z}^{\sigma \sigma'} = \sigma \im^\ell \sqrt{4\pi(2 \ell +1)} \delta_{m_z \sigma} \delta_{\sigma\sigma'}$ are the PW coefficients and $ \bm{\Psi}^{ \sigma'}_{\ell m_z}$ denote the vector spherical wavefunctions (VSWFs)~\cite{olmos2019enhanced}.  
Let us recall that these VSWFs are simultaneous eigenvectors of the square of the total angular momentum, ${J}^2$, the $z$ component of the total angular momentum, $J_z$~\cite{edmonds2016angular}, and the helicity operator for monochromatic waves, $\boldsymbol{\Lambda}$~\cite{fernandez2013electromagnetic}, with eigenvalues $\ell(\ell+1)$, $m_z$, and $\sigma$, respectively.

At this point, let us consider a homogeneous dielectric sphere of radius $a$  and  refractive index contrast ${{\rm{m}}}$ centred at the origin ($r = 0$). As the chosen system is rotationally symmetric around the OZ axis, $m_z$ is preserved in the scattering process, while the conservation of helicity, $\sigma$, crucially depends on the number of multipoles involved in the scattering process~\cite{olmos2020unveiling}. Due to a fundamental property of the Mie coefficients~\cite{PhysRevLett.125.073205,olmos2020unveiling},  preservation of helicity can only occur for lossless \emph{pure-multipolar regions}, i.e.,  spectral regimes that can be fully described by just one multipolar order $\ell$. This fact restricts duality symmetry restoration to small dipolar particles under PW illumination. 
\begin{figure}[t!]
    \centering
    \includegraphics[width=\columnwidth]{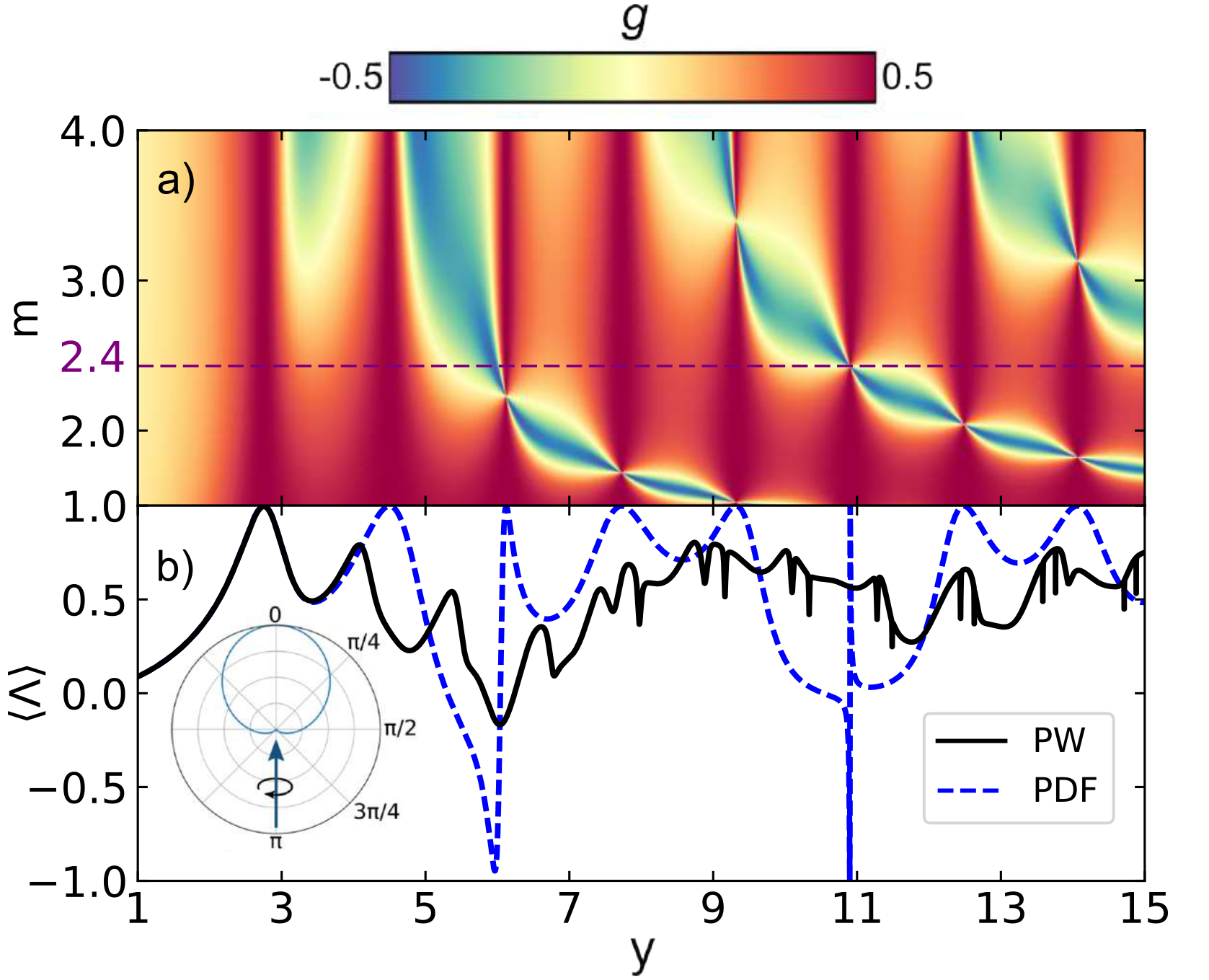}
    \caption{(a) $g$-parameter as a function of the refractive index contrast ${{{\rm{m}}}}$ and $y= {{{\rm{m}}}}ka =  {{{\rm{m}}}} x$ size parameter in the dipolar regime. (b) Helicity under both PDF and PW for a sphere  of  ${{\rm{m}}} = 2.4$. The inset illustrates the zero optical backscattering condition when $\langle \Lambda \rangle  = +1$.}
    \label{figurina_1}
\end{figure}
\begin{figure}[t!]
    \centering
    \includegraphics[width=\columnwidth]{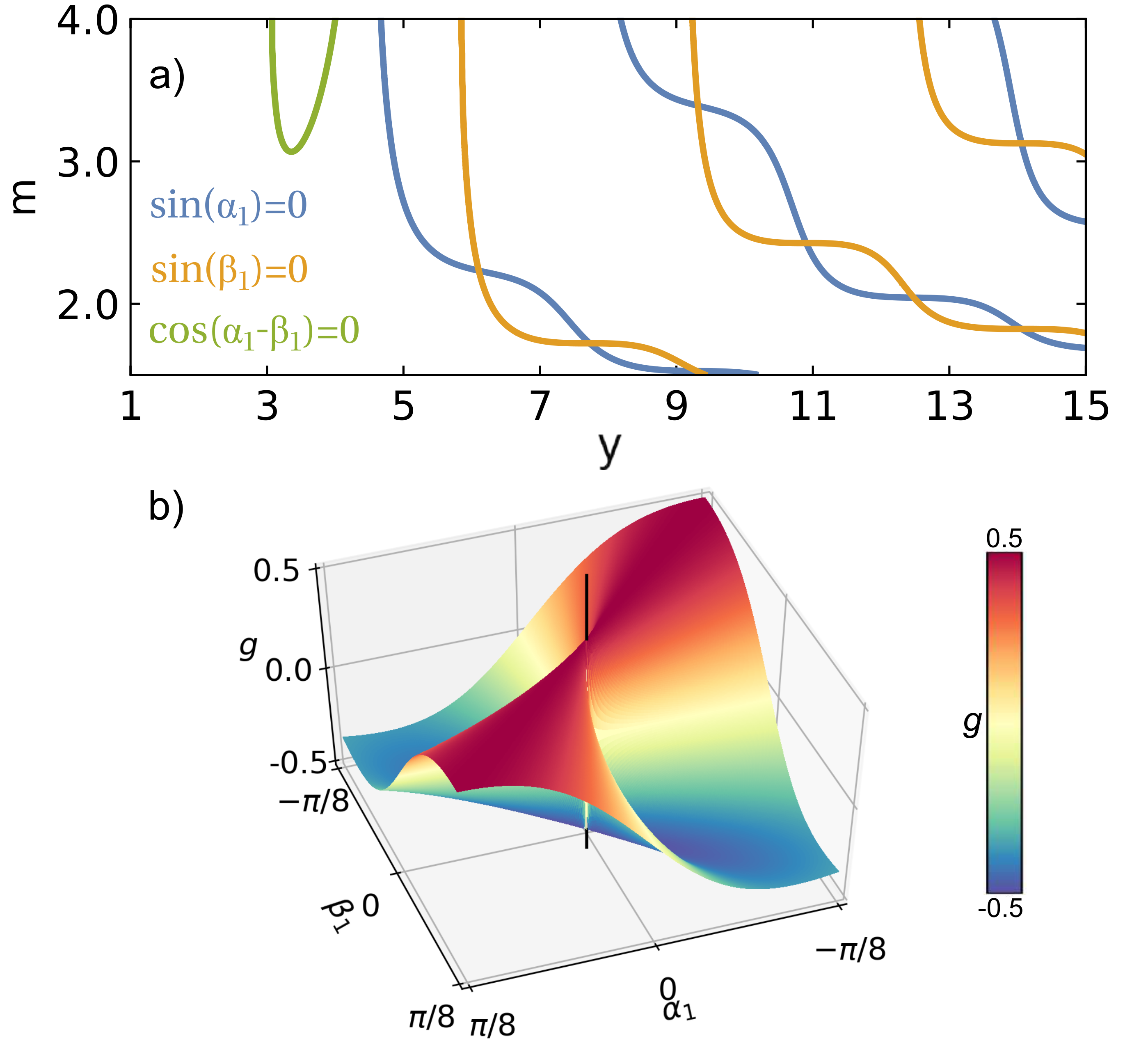}
    \caption{(a) Zeros of the $g$-parameter given by both the null amplitude of the electric and magnetic Mie coefficients (blue and orange) and by  the interference of these (green). (b) $g$-parameter in the vicinity of the Kerker anapole as a function of the dipolar  phase-shifts. }
    \label{figurina_2}
\end{figure}
To address this issue, we propose for the incident illumination, $\E^{\rm{i}}$, the use of a PDF, ($\E^{\rm{i}} \propto \boldsymbol{\Psi}_{1\sigma}^{\sigma}$), which presents a well-defined ${J}^2, $ $J_z$, and $\boldsymbol{\Lambda}$~with eigenvalues $\ell =1, m_z= \sigma = \pm 1$, respectively. As shown in Ref.~\cite{olmos2019sectoral},  this is a sectoral and propagating illumination that, by construction, only excites electric and magnetic dipoles. Notice that this phenomenon stems from the fact that, for particles placed at the focus, only the multipoles that are present in the incident beam can be excited~\cite{wozniak2015selective, zambrana2018tailoring}. To get some insight into the relevance of illuminating with a PDF, let  us consider the $g$-parameter, which encodes the  optical response of the particle via the interference between the electric and magnetic Mie coefficients~\cite{hulst1957light}. As previously mentioned,  a PDF can induce only a dipolar response, and hence, the $g$-parameter, which is constructed from the scattered Poynting vector,
\begin{equation}
g =  \langle \cos \theta \rangle = \frac{\int_\Omega  \left(\bf{S}_{\rm{s}} \cdot \bm{\hat{r}} \right) \cos \theta d\Omega}{\int_\Omega  \left(\bf{S}_{\rm{s}} \cdot \bm{\hat{r}} \right) d\Omega},
\end{equation}
is strictly identical to the expression derived from small  dipolar particles under PW illumination~\citep{olmos2020optimal}. In both scenarios, the $g$-parameter reads as 
\begin{equation}\label{eq_g}
g = \frac{\text{Re}\left\{a_1b_1^*\right\} } {  |a_1|^2+|b_1|^2} = \frac{\sin \alpha_1 \sin \beta_1 \cos \left( \alpha_1  - \beta_1 \right) }{\sin^2 \alpha_1 + \sin^2 \beta_1 },
\end{equation}
where $a_1 = \im \sin \alpha_1 e^{-\im \alpha_1}$ and $b_1 = \im \sin \beta_1 e^{-\im \beta_1}$ denote  the electric and magnetic Mie coefficients, respectively~\cite{hulst1957light}:
\begin{equation}\label{Mie_elec}
a_\ell= \frac{{{{\rm{m}}}} S_\ell({{{\rm{m}}}}x)S'_\ell(x) - S_\ell(x) S'_\ell({{{\rm{m}}}}x)}{{{{\rm{m}}}} S_\ell({{{\rm{m}}}}x)C'_\ell(x) - C_\ell(x) S'_\ell({{{\rm{m}}}}x)},
\end{equation}
and
\begin{equation}\label{Mie_mag}
b_\ell= \frac{S_\ell({{{\rm{m}}}}x)S'_\ell(x) - {{{\rm{m}}}} S_\ell(x) S'_\ell({{{\rm{m}}}}x)}{ S_\ell({{{\rm{m}}}}x)C'_\ell(x) - {{{\rm{m}}}} C_\ell(x) S'_\ell({{{\rm{m}}}}x)}.
\end{equation}
Here $S_\ell(z) = \sqrt{\frac{\pi z }{2}} J_{\ell+\frac{1}{2}}(z)$ and $C_\ell(z) = \sqrt{\frac{\pi z }{2}} H_{\ell +\frac{1}{2}}(z)$ denote the Riccati-Bessel functions.
Moreover, the  helicity and the ratio between the back and forward scattering are linked and fully determined by the $g$-parameter under a dipolar excitation~\cite{olmos2019asymmetry}, namely, 
\begin{align} \label{pure_back_asy}
\frac{\sigma_{\rm{b}}}{\sigma_{\rm{f}}} = \frac{1 -\langle \Lambda \rangle}{1 +\langle \Lambda \rangle}, && \text{with} &&
{{\langle \Lambda \rangle}}  = 2g.
\end{align}
Notice that these relations, first brought to the physical scene for small dipolar particles under PW illumination~\cite{olmos2019asymmetry}, can now be extended to arbitrary sized dielectric spheres illuminated by a PDF.
As a matter of fact, a map of $g$ in the dipolar regime for a dielectric sphere as a function of the refractive contrast index ${{\rm{m}}}$ and the size parameter $y= {{{\rm{m}}}} x = {{{\rm{m}}}} ka$, see Fig.~\ref{figurina_1}a, reveals the existence of successive first  Kerker conditions  even for low refractive index dielectric spheres. These first Kerker conditions give rise to $\langle \Lambda \rangle =2g= +1$ and, hence, to the restoration of the electromagnetic duality symmetry under PDF illumination. This phenomenon  can be inferred from Fig.~\ref{figurina_1}b, where the zero optical backscattering is also presented. 
Notice that under PW illumination the conservation of helicity requires small dipolar particles, e.g., $y<4$.

\begin{figure}[t!]
    \centering
    \includegraphics[width=\columnwidth]{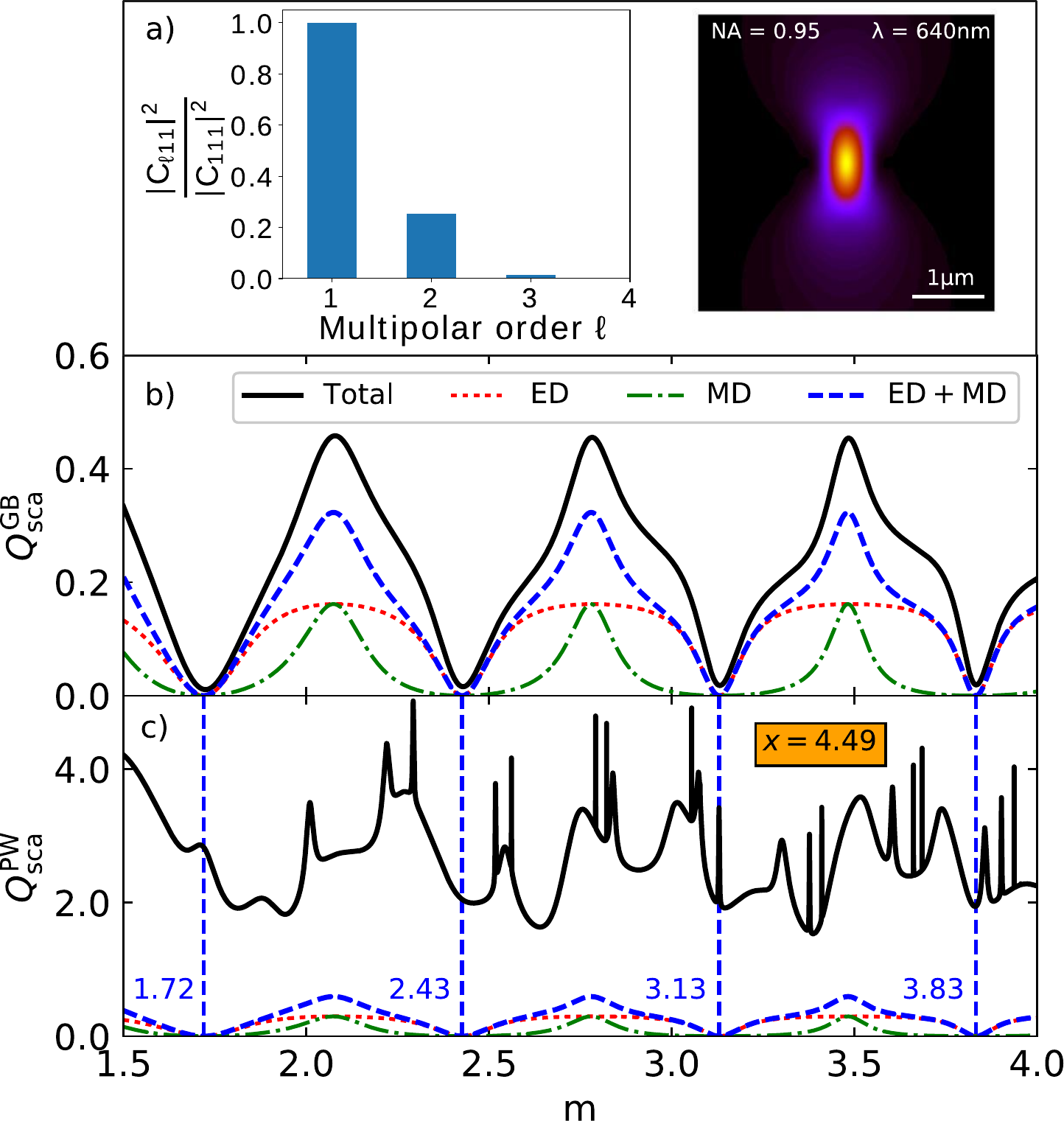}
    \caption{(a) Relative weights of each multipolar order with respect to the dipolar contribution and sketch of the  tightly focused GB. Scattering efficiency from  dielectric spheres  versus the contrast index ${{{\rm{m}}}}$ for a fixed size parameter $x = 4.49$ under the GB illumination (b) and PW illumination (c). The contributions of the electric and magnetic dipoles  to the total scattering efficiency are presented.  }
    \label{figurina_3}
\end{figure}

Moreover, an interesting phenomenon that is unveiled under PDF illumination is the emergence of pure electric (or magnetic) scattering regimes  for $\sin \beta_1 = 0$ (or $\sin \alpha_1 = 0$). Note that both conditions give rise to  $g = 0$, according to \eqref{eq_g}. In order to infer each scattering regime in Fig.~\ref{figurina_1}a, we distinguish in Fig.~\ref{figurina_2}a the three solutions that lead to $g=0$. These correspond to the destructive interference of the electric and magnetic Mie coefficients, namely, $\alpha_1 = \beta_1 \pm\pi/2$ (green line) or to the abovementioned pure electric ($\sin \beta_1 = 0$) or magnetic ($\sin \alpha_1 = 0$)  scattering regimes in orange and blue lines, respectively.
Particularly, when these latter  two lines cross each other, zero scattering points appear under dipolar excitation, i.e., the scattering cross section $\sigma_{\rm{sca}} \propto \left( \sin^2 \alpha_1 + \sin^2 \beta_1 \right) = 0$, and, hence, the sphere is effectively invisible to the incoming light.
These singularities, which are visible in Fig.~\ref{figurina_1}, have been predicted in a recent theoretical work under the name of \emph{hybrid anapole modes}~\cite{luk2017hybrid}. 
In that work, the solutions of the so-called hybrid anapoles are found by solving transcendental equations that require numerical methods. However,  it is straightforward to notice that these anapoles arise analytically by imposing the first Kerker condition with zero scattering, i.e., $S'_\ell({{\rm{m}}}x) = 0$ with $S'_\ell(x) =0$ or $S_\ell({{\rm{m}}}x) = 0$ with $S_\ell(x) =0$. Note that both conditions lead to $|a_1| = |b_1| = 0$, according to \eqref{Mie_elec} and \eqref{Mie_mag}.
To get a deeper understanding of these non-radiating anapoles, we have depicted the $g$-parameter as a function of the dipolar electric and magnetic scattering phase-shifts (see Eq. \eqref{eq_g}) in Fig.~\ref{figurina_2}b. As it can be inferred, the anapole emerges, under PDF illumination, when the condition $\alpha_1 = \beta_1 = 0 $ is met (see the vertical black line). Strikingly,  this optical invisibility condition presents a direct connection with both Kerker conditions and light transport phenomena in spheres: at the first Kerker condition ($\alpha_1 = \beta_1$) the $g$-parameter is maximized, $g=0.5$ (intense red colour) and, then, the zero optical backscattering condition is met. On the other hand, when $\alpha_1 = -\beta_1$, the generalized second Kerker condition (GSKC) is satisfied (intense blue colour) and, then, the optimal backward scattering condition is guaranteed~\citep{olmos2020optimal}. Notice that achieving these optical responses (either first Kerker condition or GSKC) in the vicinity of these anapoles not only changes the directionality of the scattered light drastically but also the transport mean free path~\cite{jones2015optical, gomez2012negative} and the dipolar optical forces~\cite{marago2013optical}, as  $g$ can flip from $g = 0.5$ (duality) to $g \sim -0.5$ (nearly anti-duality~\cite{ zambrana2013duality}) within a small perturbation of the scattering phase-shifts. Figure.~\ref{figurina_1}b) summarizes the physics behind these phenomena at $y \sim 11$ where we depict the abrupt change of the scattered helicity of a  lossless sphere with ${{{\rm{m}}}} = 2.4$. 
In this line, it is crucial to notice that these non-radiating anapoles cannot be found for lossy spheres since $a_\ell \neq b_\ell~ \forall \ell$ in the presence of absorption or gain, as mathematically demonstrated in Ref~\cite{PhysRevLett.125.073205}. On physical grounds,  non-radiating anapoles cannot emerge for lossy systems such as plasmonic particles or silicon spheres in the visible spectral range~\cite{olmos2019role} since an object that absorbs must radiate. For all the reasons mentioned above, we honestly believe that "Kerker anapole" is a name that more precisely describes the origin of this type of anapoles.

\begin{figure}[t!]
    \centering
    \includegraphics[width=\columnwidth]{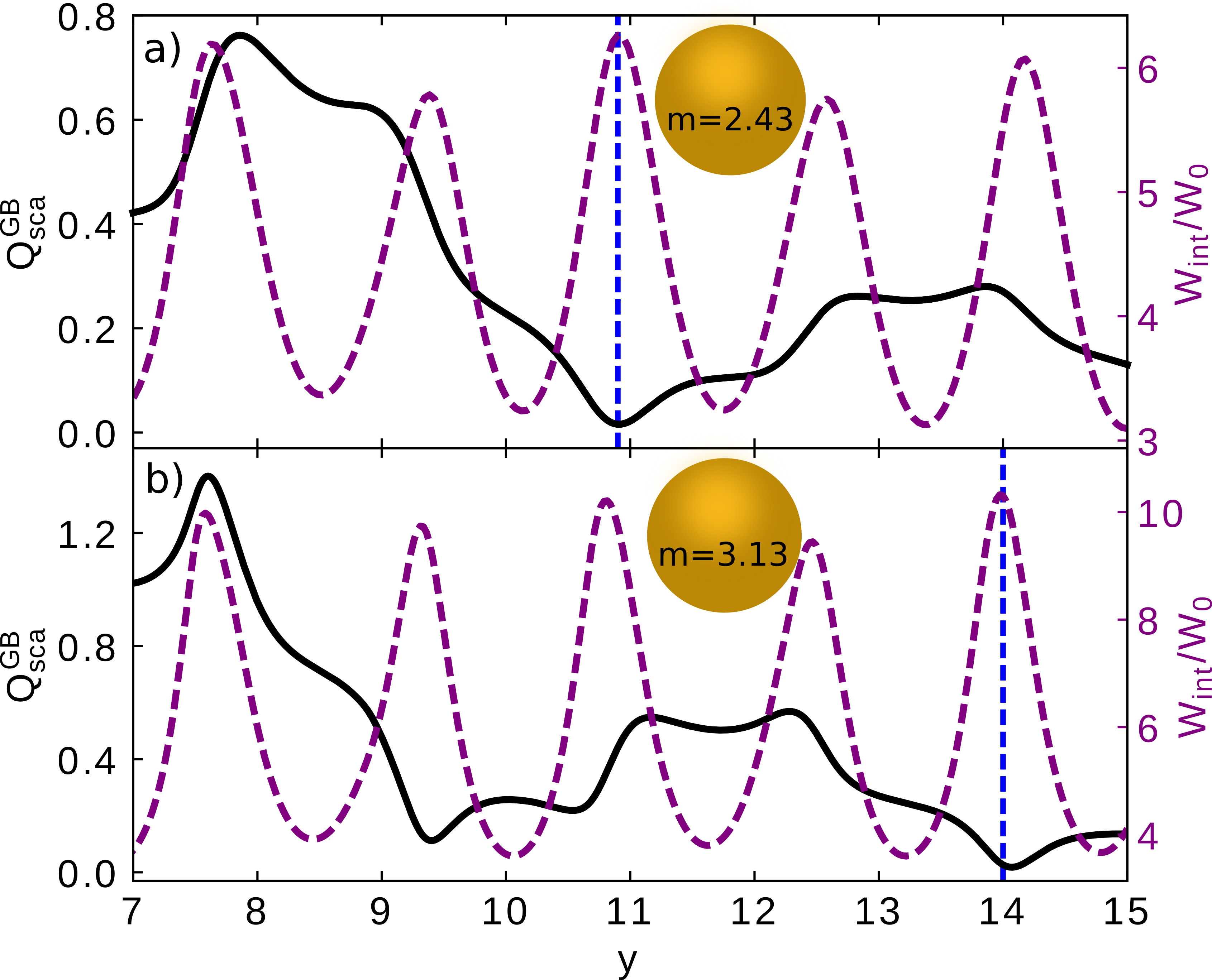}
    \caption{Scattering efficiency (black) and  internal electromagnetic energy (purple) under GB illumination versus the $y$ size parameter for two refractive indexes: (a) ${{{{\rm{m}}}}}= 2.43$ and (b) ${{{{\rm{m}}}}} = 3.13$. The vertical dashed blue lines illustrate the maximum of the confined energy, which lies in the vicinity of the Kerker anapole.}
    \label{fig_ultima}
\end{figure}

Hitherto, we have discussed the optical properties of homogeneous dielectric spheres under a PDF excitation. However, even though these sectoral and propagating beams fulfill Maxwell's equations~\citep{olmos2019sectoral}, its experimental implementation is not standard yet~\citep{zaza2019size, huang2020surface, aibara2020dynamic}. 
In this regard, we alternatively propose the use of tightly-focused Gaussian beams~\cite{allen1992orbital, gori1994flattened, mojarad2008plasmon, orlov2012analytical, zambrana2012excitation} whose multipolar decomposition has been extensively analyzed.  
Specifically, we apply the method presented in Ref.~\citep{zambrana2012excitation}, in which the multipolar decomposition is determined by the beam coefficients $C^{\sigma}_{\ell m_z}$. The independent control of both angular momentum and polarization, allows for the generation of a GB with well-defined $m_z = \sigma = 1$. In this case, the relative weight of each multipolar order with respect to the dipolar order  can be highly minimized when the beam is tightly focused at the center of the object. As a matter of fact,  by using commercial data from a  high numerical aperture, e.g. NA $= 0.95$ and $f=1.8$ mm, and fixing the optimal ratio between focal length and beam waist, $f/\omega_0 \sim 0.9$, the relative weights can be straightforwardly obtained. These are shown in Fig.~\ref{figurina_3}a, where the tightly focused GB is also presented.  According to the abovementioned, the scattering efficiency of a sphere illuminated by a tightly focused Gaussian beam with a well-defined helicity is~\cite{zambrana2012excitation}
\begin{equation}\label{eq_Qsca_GB}
Q^{\rm{GB}}_{\rm{{sca}}} =  \sum_\ell \frac{\left(2\ell +1 \right)}{x^2} |C_{\ell 1 1}|^2 \left( \sin^2 \alpha_\ell  + \sin^2 \beta_\ell  \right)\;,
\end{equation}
where the value of the beam shape function $C_{\ell11}$ heavily decreases as the multipolar order $\ell$ is increased.
By using the computed relative weights shown in Fig.~\ref{figurina_3}a, we calculate in Fig.~\ref{figurina_3}b the scattering efficiency as a function of the contrast index ${{\rm{m}}}$ for a fixed size parameter of $x=4.49$. As it can be inferred,  the scattering efficiency practically vanishes for several refractive indexes, namely,  $Q^{\rm{GB}}_{\rm{{sca}}} \sim 9 \cdot 10^{-3}$, for ${{{\rm{m}}}} = 1.72, 2.43, 3.13, 3.83$. These are examples of non-radiating Kerker anapoles that give rise to the singularities of the  $g$-parameter under PDF illumination, according to  Fig~\ref{figurina_1}a. However, under PW illumination (see Fig.~\ref{figurina_3}c), Kerker anapoles cannot be characterized as non-radiating sources since higher multipolar orders predominantly contribute to the optical response of the spheres.
As a further verification of the existence of non-radiating sources under the considered tightly-focused GB, we also compute the internal electromagnetic energy normalized with respect to the GB energy in the absence of particle. The internal electromagnetic energy is calculated from the internal fields~\cite{zambrana2012excitation}, $\E^{\rm{int}}$ and $\H^{\rm{int}}$, using the internal Mie coefficients~\cite{hulst1957light}. Therefore, the ratio between internal, $W^{\rm{int}}$, and incident, $W^{\rm{i}}$, energies in the particle's volume, $V$, reads as
\begin{equation}
   \frac{W^{\rm{int}}}{W^{\rm{i}}} = \epsilon_{\rm{r}} \frac{\int (|\E^{\rm{int}}|^2 + Z^2 |\H^{\rm{int}}|^2)\ dV}{ \int (|\E^{\rm{i}}|^2 + Z_0^2 |\H^{\rm{i}}|^2)\ dV},
\end{equation}
where $\epsilon_{\rm{r}} = {{{{\rm{m}}}}}_{\rm{p}}^2$,  $Z=\sqrt{\mu/\epsilon}=(1/{{{{\rm{m}}}}}_{\rm{p}})\sqrt{\mu_0/\epsilon_0}=Z_0/{{{{\rm{m}}}}}_{\rm{p}}$, and ${{{{\rm{m}}}}}_{\rm{p}}$ is refractive index of the particle.
As it can be concluded from Fig.~\ref{fig_ultima}, the scattering efficiency is practically suppressed at the Kerker anapoles while the normalized internal electromagnetic is considerably enhanced. This phenomenon confirms the existence of non-radiating induced sources in the depicted spheres. Importantly, these are natural materials in the visible spectral range: the Kerker anapole in Fig.~\ref{fig_ultima}a arises for a diamond sphere embedded in air with a radius of $380$ nm and $\lambda_0 = 532$ nm. The other Kerker anapole in Fig.~\ref{fig_ultima}b corresponds to an aluminium arsenide (AlAs) sphere embedded in air with a radius of $430$ nm and $\lambda_0 = 604$ nm.

In conclusion, we have shown that dipolar spectral regimes can be unveiled under PDF illumination, regardless of both the size of the sphere and its refractive index. As a result, we have demonstrated that unexplored Kerker conditions can lead to backward-to-forward asymmetric scattering, or equivalently, from dual to anti-dual behaviour even for microsized spheres. Moreover, we have shown that non-radiating Kerker anapoles emerge for several microsized dielectric spheres behaving dipolarly, opening new insights into the optical invisibility. Finally, we have also demonstrated that Gaussian beams tightly-focused by commercial microscopic objectives could mimic the scattering features produced by these PDFs. These findings could drive the implementation of new experiments beyond the actual physical picture, which is mostly restricted to HRI materials in the limit of small particle.

The authors dedicate this work to the memory of their beloved colleague and friend, Prof. Juan José Sáenz (Mole), who passed away on March 22, 2020.

C.S-F. acknowledges funding by the
Spanish Ministerio de Ciencia, Innovación y Universidades
(MICINN) (Project No. FIS2017-82804-P). M. M. acknowledges financial support from the MICIU through the FPI PhD Fellowship FIS-2017-87363-P. N.dS. acknowledges support from the Spanish Ministerio de Ciencia e Innovacion (grant number PID2019-109905GA-C22).  X.Z-P. acknowledges
funding from the European Union’s Horizon 2020
research and innovation programme under the Marie Sklodowska-Curie grant agreement No 795838.
. J.O-T. acknowledges funding by the Basque Government (Project PI-2016-1-0041 and PhD Fellowship PRE-2018-
2-0252).

\bibliography{New_era_18_06_2019}

\end{document}